\documentclass{epl}
\title {Observation of narrow fluorescence from doubly-driven four-level atoms 
at room temperature} 
\author{Uday Kumar  Khan, Jimmy Sebastian, N. Kamaraju, 
Andal  Narayanan,\\ R. Srinivasan and Hema Ramachandran}
\institute{Raman Research Institute,  Sadashivnagar, Bangalore, INDIA 560 080}
\begin{document}
\maketitle
\begin {abstract} 

\indent Unusually narrow fluorescence peaks are seen from 
$^{85}Rb$ atoms  under the action of two driving laser fields that are in a
three-dimensional molasses configuration. One of the lasers is held at a fixed 
detuning from the "cooling" transition, while the other is scanned across the 
"repumping" transitions. The fluorescence 
peaks are split into symmetric pairs, with the separation 
within a pair increasing
with the detuning of the cooling laser. For large detunings, 
additional small peaks
are seen. A simple model is proposed to explain these experimental observations. 
An exhaustive density-matrix calculation  of the four level system is presented in
a companion paper. \\
\end {abstract} 

\section {Introduction} 
\indent Cooling of alkali atoms  from room temperature
to a few hundred microkelvin   using near-resonant light   
has now become a standard technique for obtaining cold atoms.
An area of current interest is the response of the cold atoms
to multiple laser fields.  The response
may be incoherent, giving rise, for example,
to sequential  transitions, or coherent, 
resulting  in quantum interference effects
like  Electromagnetically Induced Transparency (EIT) \cite{EIT},
Coherent Population Trapping,
Lasing Without Inversion (LWI) \cite{LWI} etc. Underlying most of these
phenomena is the fundamental response of an  atom to a strong pump beam,
  in the form of ac Stark splitting of the  electronic levels.  
Also known as   Autler-Townes  dressed states \cite{AT}, these manifest themselves 
as  modifications to  the absorption spectrum. A large volume of 
theoretical work on the  absorption and emission
from  three level atoms under the action of  pump and probe 
fields exists \cite{pupo} and  several 
experimental  studies of the absorption spectra that verify
 some of the predictions have been reported \cite{three}. 
However, there are very few \cite{four} investigations of the  fluorescence
from  four level systems under the action of  several fields.

\indent We report in this paper, fluorescence measurements from a doubly driven
four level system. The alkali atom $^{85}Rb$ under the action of
two laser fields, the "cooling" 
and "repumper" fields, in a three dimensional optical
molasses configuration is studied.
Unusual features are seen in the fluorescence when  one of the lasers, the
"cooling" laser 
is held at fixed frequency, and the other scanned.
We discuss in simple physical terms, a  mechanism that gives some insight into the 
experimental observations. Detailed theoretical analysis in terms of a four-level
 density matrix is presented in a companion paper \cite{compa}. 
\section {Experiment}
The energy levels of
$^{85}Rb$ (Fig. 1) show  that the closed 
transition $5S_{1/2}F=3 -> 5P_{3/2}F'=4$ \footnote { Hyperfine levels F 
of the state $5S_{1/2}$ will be denoted unprimed and those of
$5P_{3/2}$ as F'  or primed. } is appropriate for cooling.
As is well known, even when the cooling laser is resonant with the cooling transition,
 due to their proximity adjacent hyperfine levels also get populated. These levels can 
then decay, in addition to the F=3 level, to the other ground level F=2 also, 
 thus depleting the cooling cycle in a few absorption-emission cycles. This necessitates
the use of a repumper laser, tuned to the F=2 $->$ F'= 3 or 2, transition
which serves to bring back atoms to the F=3 state, and thus into the cooling cycle.\\
\indent  In our experiment we use two external cavity diode lasers, locked to the 
F = 3 $->$ F' = 4  and  F = 2 $ ->$ F'=3 transitions for  the
 cooling and the repumper beams.
The beams from each of the two lasers were split into 
four parts, one of which was sent to a Doppler-free saturation absorption setup in a vapour
cell that was used to determine the instantaneous frequency of the laser when it was ramped 
or was  used for locking the frequency. The remaining three beams from each laser 
 were incident into the chamber from the 
the +X, +Y and +Z directions, and were retroreflected, to form three pairs of
counterpropagating, mutually orthogonal beams, that intersected at 
the centre of the chamber. 
\begin{figure}
\onefigure{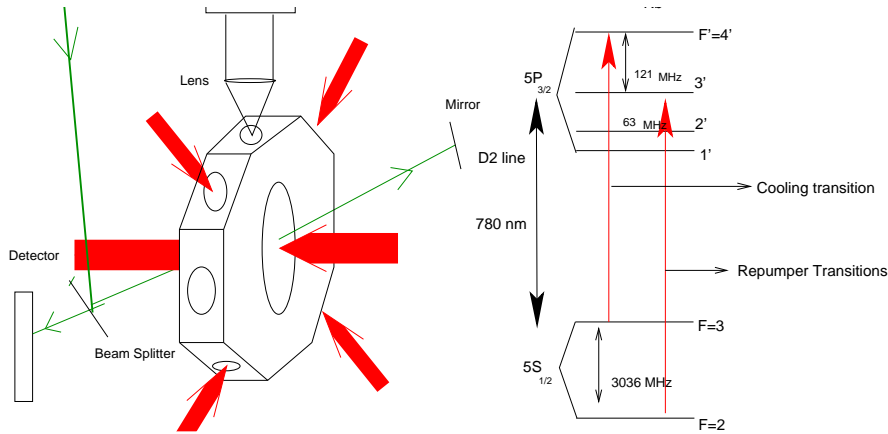}
\caption{The experimental setup :
the thick red lines are the cooling and repumper beams, the green line
the absorption probe.
The energy level diagram of $^{85}Rb$ is given on the
right.
}
\end{figure}
The experimental chamber (Fig.1), made of stainless steel,
was evacuated to $2$x$10^{-9}$Torr using a 40l/s ion pump. The eight faces,
of the octagonal chamber  and the two end caps had
one viewport each. Rb atoms were obtained from  a getter source. 
While  in
laser cooling experiments, the 
repumper beam is usually sent in  one direction only, 
in our case the repumper beams were sent in all three directions and  
retroreflected. This has given rise to unexpected features, as
 will be discussed shortly. The diameter of the beams was
 about 1 cm, and thus the molasses region was a cube of side 1cm. 
Each cooling beam was of intensity $1$mW/cm$^2$, and the repumper $0.1$mW/cm$^2$.
A femtowatt detector was placed at one of the viewports  and a  lens system imaged 
 the  fluorescent light from  the centre of the chamber onto the detector.  
We also studied the absorption of a weak probe, in the  Doppler-free pump-probe
 configuration shown in Fig.1.  

\indent Fluorescence from the atoms was studied when they were 
driven by two fields, the cooling and the repumper laser fields. 
The cooling laser was locked at various detunings $\delta_c$ in the range 
+120MHz (blue) to -200MHz (red) with respect to the cooling transition.
The detunings thus spanned a region  much greater than the full 
hyperfine manifold  F=3 $-> $ F'.
 At each detuning the repumper laser was scanned very slowly (0.05Hz)
across its  hyperfine manifold, i.e.,  $F=2 ->  F'$.\\
\indent We remark here that cooling of atoms occurs when the cooling laser is detuned to the
red within a few linewidths of the cooling transition. In our experiment, however,
the detunings are often quite large, and in many cases to the blue of the cooling transition.
Even though the "cooling" beam does not cool, nor does the "molasses" provide a retarding
force, we retain terms like  "cooling" and "molasses".\\
\begin{figure}
\onefigure{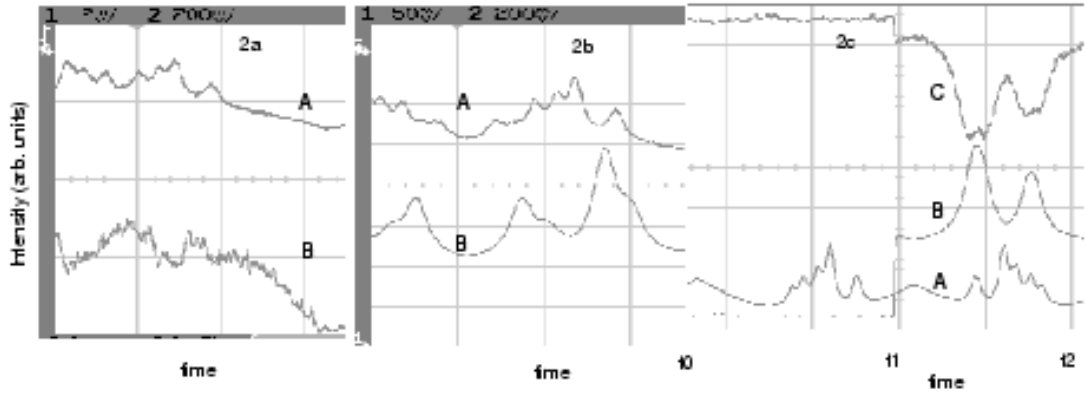}
\caption{
(2a) : B - Fluorescence from the molasses region when only the repumper is incident.
(2b) : B - Fluorescence from the molasses region when both repumper and cooling beams are 
present.
(2c) : B - Fluorescence from molasses region. 
       C - Absorption of a probe (cooling laser) transiting the molasses.
Trace A in all three cases is the saturation absorption signal of the frequency scanned repumper
and is a measure of the instantaneous repumper frequency. }
\end{figure}
\indent When only the repumper beam is allowed into the chamber 
the fluorescence 
is weak (~2mV)(Fig.2a), as opposed to a stronger fluorescence (~50mV) 
when the cooling beam is also
incident on the atoms (Fig. 2b).
All the results presented here, have fluorescence in the 50 mV range, 
indicating that it is predominantly from the cooling 
transition ($4' -> 3$). We note 
that while $F=2 -> F' = 1,2,3$ are allowed transitions,
only $F' = 2,3$ serve to
repump  atoms back into the cooling cycle. As we see enhanced fluorescence 
 only when the repumper
laser is close to either of these transitions (Fig.2b), it confirms that it is predominantly 
the cooling transition that contributes to the fluorescence. 
Fig. 2c  illustrates  an interesting feature. 
  Trace A shows the saturated absorption spectrum of the repumper laser 
that is periodically scanned across $F=2 -> F'$ manifold. Trace B is the fluorescence 
signal from the molasses and trace C the absorption of a weak probe derived from the cooling 
laser that is locked to the $F=3 -> F'=4$ transition. For  the first half of the trace,  
the repumper laser to the  molasses is blocked. During this period, the 
fluorescence and absorption signals  are unchanged in time, which 
is  as they should be, as the cooling laser is locked.  
In the later half, the cooling laser continued to be locked, but the repumper,
with its frequency being periodically scanned,  was
 allowed into the molasses region. The 
presence of the repumper  dramatically alters the absorption of the cooling laser, as is 
seen from trace C. Further, as trace B is roughly similar (though inverted)
to trace C, it is reasonable to conclude that the light absorbed 
from the cooling beam
by the atom is spontaneously emitted in random directions, and
forms the fluorescence signal. \\
\indent The dependence of the fluorescence signal on the
detunings of the two driving fields is shown in Fig. 3a ($\delta_c<0)$ and
3b ($\delta_c>0)$. In both figures, trace A shows the saturation 
absorption spectrum of the repumper as it is scanned across $F=2 ->$ F', providing a 
measure of the instantaneous repumper frequency. The remaining traces display the 
instantaneous fluorescence intensity collected from the 
molasses region, for various detunings of the cooling laser from the F=3 $-> $F'=4 
cooling transition. 
Trace B in both figures shows the fluorescence from the molasses with 
the cooling laser  on resonance. The fluorescence shows two maxima,
( labelled {\it'a'} and {\it 'b'} ), 
corresponding to the two repumping  transitions. For small detunings of the cooling laser
(Traces 3a(C) and 3b(C) ), narrow dips appear in the fluorescence signal, splitting the peaks
{\it a} and {\it b}.  These are centred at 
the points where the repumper laser is exactly resonant with the repumping 
transitions. Each of the  peaks {\it a} and {\it b}  split into a pair of peaks,
 labelled {\it $a_+$}, {\it $a_-$} and  {\it $b_+$},{\it $b_-$},  symmterically displaced
from the original peaks, a and b. 
As the cooling laser is progressively detuned, each pair  {\it $a_+$}, 
{\it $a_-$} and  {\it $b_+$} , {\it $b_-$} separate further (Traces D-F in 3a and 3b )
 resulting in 
four distinct peaks  in the fluorescence signal.  The width of each peak remains 
fairly constant at  30MHz.  In several cases,
 there appear some very weak peaks, indicated by arrows.
 These peaks are well resolved
 only for large detunings of the cooling laser. 
In Fig. 3c, the frequency shift
of each of the four peaks with respect to the unsplit
repumper transition ($F=2 -> F'=3$) is plotted as a 
function of the detuning of the cooling laser $\delta_c$. 
The peak positions fall on 
straight lines with slopes $\pm1$; the two lines corresponding to the + and - peaks of 
each transition intersect when the detuning of the cooling laser is zero. 
The intercepts
are separated by $\approx 63$ MHz, 
which  corresponds to the hyperfine interval between the 
upper levels $F' = 2$ and 3 of the repumping transition. 
\begin{figure}
\onefigure{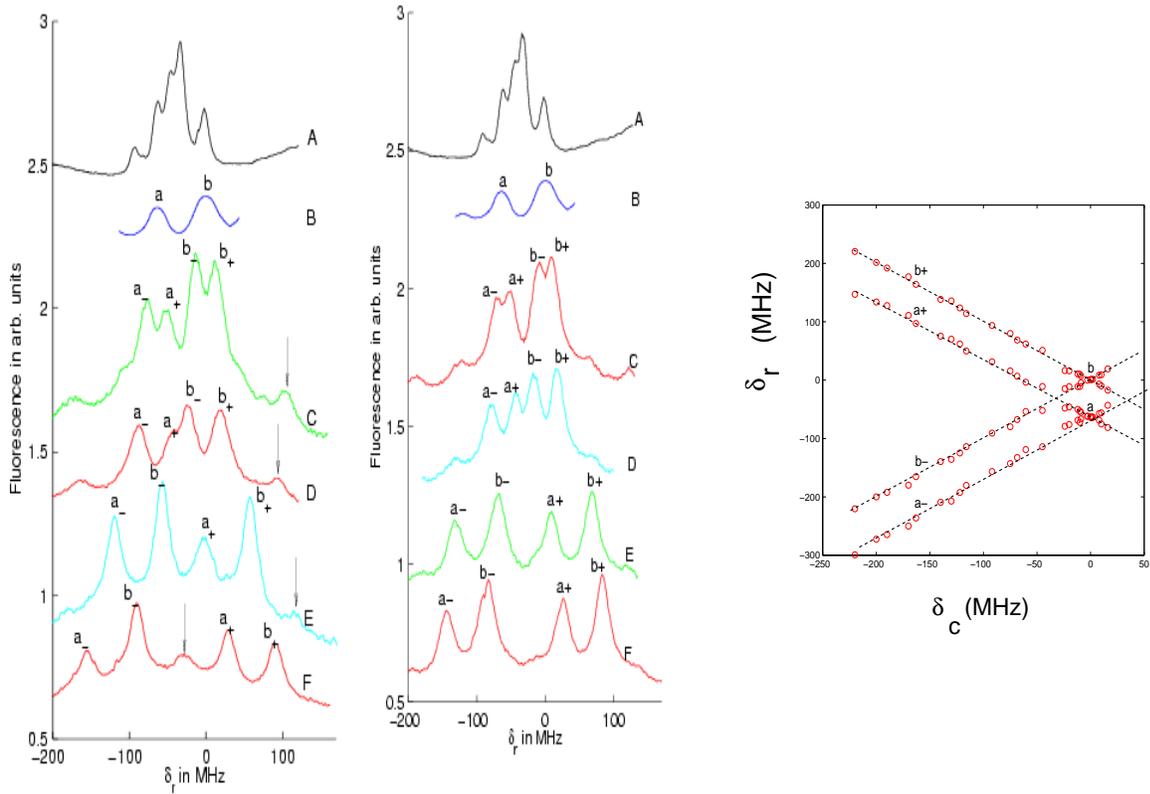}
\caption{
Fluorescence signal from the molasses for (3a)  
red detuning  and (3b) blue detuning of
the cooling laser as a function of the instantaneous repumper frequency.The uppermost trace gives
the saturation absorption spectrum of the repumper laser.
Traces B-F correspond to $\delta_c = 0, -12, -22, -60$ and -92 MHz in
(3a) and to $0, 10, 20, 70$ and 90 MHz in (3b).
3c : The positions of the fluorescence peaks as a function of the detuning of the
cooling laser. The dotted lines have slopes $\pm 1$ and are drawn as a guide to the eye.
$\delta_r =0$ corresponds to the $2 -> 3'$ transition. 
}
\end{figure}
\begin{figure}
\onefigure{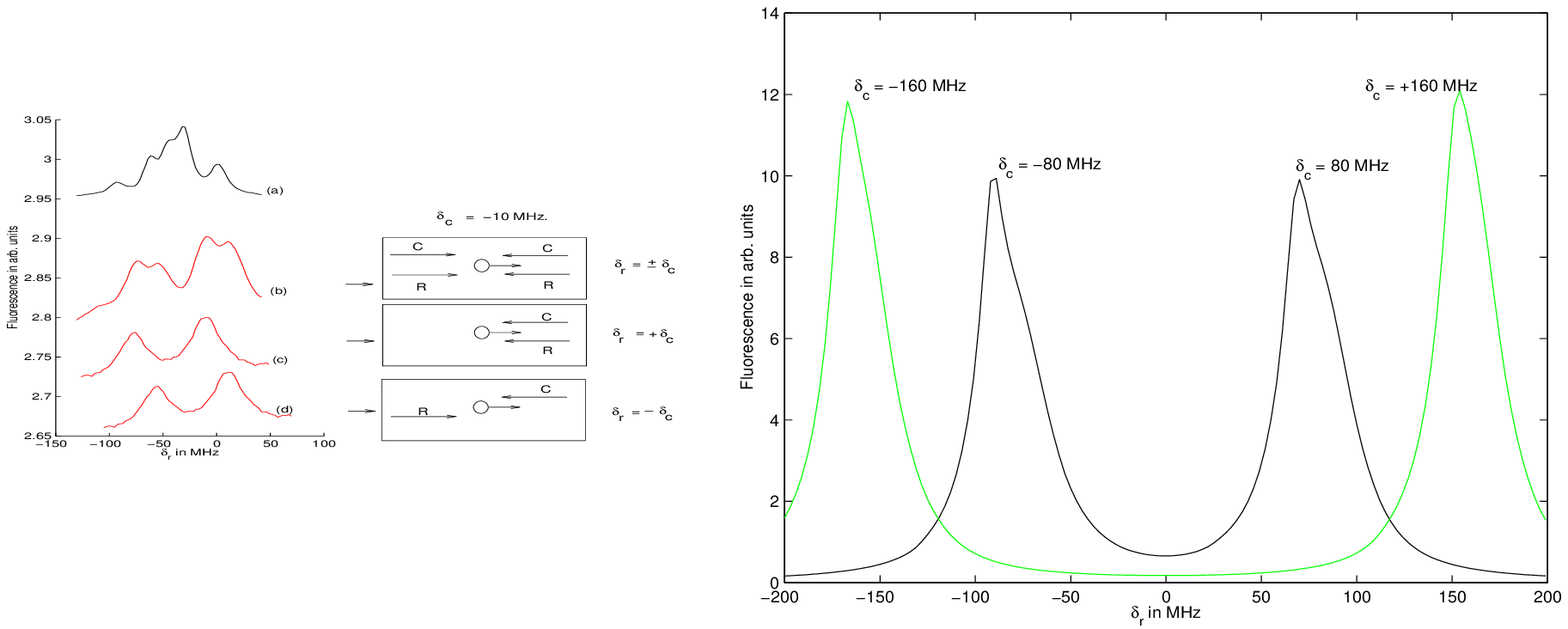}
\caption{(a): Trace a: Saturation absorption signal of the repumper. Trace b,c and d
are the fluorescence signals obtained for the beam configurations shown alongside.
(b): Fluorescence estimated using equation (4) 
for $\delta_c = -100$ MHz (green) and $\delta_c = 80$ MHz (black).
}
\end{figure}
\section {Discussion}
        Typical experiments on two and three level atoms  using multiple fields
study the absorption of a weak probe that  has counter-propagating to it 
a strong beam of the same frequency, that enables the
 selection of atoms with zero velocity along the direction   of the 
two beams. It is in this configuration that one studies narrow spectral features,
that could otherwise not have been resolved due to the Doppler shift. By the
same token, fluorescence measurements are not expected to display narrow spectral 
features as they are emissions from a collection of randomly moving atoms. 
It therefore comes as a surprise that our experiment shows dips in the fluorescence 
as narrow as 6MHz, separating  peaks with widths of 30MHz.\\
\indent It may be noted that while we are not performing any frequency
analysis of the emitted fluorescent light, we are 
studying the fluorescence  response of the atoms to the repumper laser 
that is being frequency scanned. The various spectral features occur at the same
relative positions to the repumper peaks for all slow rates of scan.

\indent
The narrow spectral features suggests a quantum phenomenon. In this
context, one could look for Autler Townes splitting, where, due to the presence of 
a strong pump laser, the upper $\footnote {lower levels do not split as J=1/2}$ levels
get split. In our case, the pump laser is the cooling laser, which  
splits the upper repumper levels ($F'$ = 2,3). This split would be directly discernible 
if one were looking for absorption to, or emission from these levels. In 
our experiment, we
are recording  predominantly, the emission due to the cooling transition 
($F'=4  \rightarrow F=3$).
It is, however,
possible for the Autler Townes splitting of the upper repumper
levels (F'=2 and 3) to be indirectly 
reflected in the fluorescence signal from  $F'=4$ level.\\ 
\indent In the absence of the cooling laser, when the repumper is scanned, 
atoms with zero velocity will  make an upward transition from 
$ F= 2 -> 3'$  when the frequency of the laser $\omega_{23'}$ is
the transition 
frequency. In the presence of the strong  cooling laser, 
the $F'=3$ level will  split, giving rise to the Autler-Townes doublet  
with displacements  $\delta_{\pm} = \delta_p /2 \pm
(\delta_p^2  + \Omega^2)^{1/2}/2 $ \cite{pupo}. $\delta_p$ refers to
the detuning of the pump (cooling) laser from the  F=3 $->$  F'=3' transition,
(i.e., $\delta_c + 2 \pi *121 MHz$), 
and $\Omega$  its Rabi frequency. 
Thus, when the repumper laser frequency is scanned about $\omega_{23'}$,
 atoms will be  transferred out of the 
 F=2  level twice, when the laser is at  $\omega_{23'} + \delta_-$
and at  $\omega_{23'} + \delta_+$. 
At both these instants, nearly all atoms will be  
transferred to the level  F=3 via the upper repumper level. These atoms can then 
participate in the "cooling" transition, resulting in an enhanced
 fluorescence thus giving an
indirect evidence of the Autler-Townes split \footnote{The upper cooling level 
(F=4) too, should experience a similar ac 
Stark splitting.
This, however, will not be seen in our measurements as we are not scanning the
cooling laser, nor are we  performing a frequency 
discrimination of the emitted fluorescence signal. }.
In our experiment, we do indeed see maximal repumping efficiency for two values 
of the repumper frequency about each repumping transition. 
However, the peak positions are symmetrically displaced about the unsplit
position which is unlike the Autler-Townes doublet, where for example,
for small $\Omega$, the shifts are 0 and $\delta_p$. 
Further, the fluorescence is narrow, and not Doppler broadened as expected.
\section{The Double Resonance Model} 
Maximum fluorescence is obtained when atoms are repumped 
from F=2 level to F=3 level, and then raised to F'=4. This requires the 
enhanced transition rates from 2$-> $3' and 3$->$ 4' or, 2$->$2' and 3$->$ 4', or
equivalently, increased absorption rates of a repumper photon and a cooling photon.
The requirement that two photons be absorbed, in turn, requires that the detuning
 of the two lasers  be compensated for by the Doppler shift. This provides a means 
of velocity selection where atoms of a particular velocity can be selected
depending on the detuning of the pump laser. This is in contrast to the  
conventional Doppler-free
saturation absorption, where only zero-velocity atoms are selected out. \\
\indent For an atom with velocity $\vec{v}$, and cooling beam of detuning $\delta_c$, 
 absorption is maximised when $\vec{k}.\vec{v} = -\delta_c$. The repumper, which serves to bring
atoms back to the cooling cycle, will be most effective for  this 
velocity class of atoms only when its detuning $\delta_r = \pm \vec{k}.\vec{v}$.
The '+' (-) sign is for the repumper beam 
counter- (co-) propagating to the cooling beam. Both possibilities exist in our 
experiment. 
If $\delta_r \ne \pm \vec{k}. \vec{v}$ repumping will not occur and  the fluorescence 
from the cooling transition  will be diminished.
This is precisely what is seen in the experiment.  
The slopes $\pm1$ in Fig.3c indicate that the maximum fluorescence occurs only when
$\delta_c = \pm \delta_r$, that is, only when  the resonance
condition is simultaneously satisfied for both the cooling and repumper beams.
 Since we have two repumper transitions we see two pairs of lines.
For $\delta_c = 0$ 
the two pairs of lines converge at $\delta_r \approx 0$ and 
$\delta_r \approx -63$MHz, this  being
the  hyperfine interval between F'=3 and  F'=2 (Fig. 3c).
On the lines of \cite{three}, this suggests a 
means of determining hyperfine intervals to good accuracy. \\
\indent To verify the validity of this model,
experiments where performed
for various configurations of co- and counter- propagation of the laser beams. 
The results are shown in Fig. 4a.
When the cooling and repumper beams
are both in the same direction ($\vec{k}$), an atom with velocity $\vec{v}$,
sees them as $\omega_c + \delta_c - \vec{k}.\vec{v}$ and 
$\omega_r + \delta_r - \vec{k}.\vec{v}$.
For the two transitions to be simultaneously on resonance, 
we require
$\delta_c = \delta_r = \vec{k}.\vec{v}$.
The experiment did indeed yield a  peak at
$\delta_r$ = $\delta_c$ only. 
For the case, where  the repumper is counterpropagating to the cooling beam,
an atom of velocity v will see  the lasers at $\omega_r + \delta_r + \vec{k}.\vec{v}$ and 
$\omega_c +\delta_c - \vec{k}.\vec{v}$, giving the condition $\delta_c = -\delta_r$ for
simultaneous resonance of the two transitions. Once again, the experiment
bore this out. When the cooling and repumper beams are 
incident in both directions, peaks appear at both $\pm \delta_c$,
confirming  that  simultaneous Doppler resonance is the 
origin of the symmetric pairs of fluorescence peaks. This, surprisingly, gives rise to 
a fluorescence spectrum of line width as narrow as 30 MHz from a 
collection of atoms of Doppler width 500MHz. 
\section{Calculation of fluorescence peaks}
\indent We now give an  analytical derivation of the fluorescence
spectra to show that narrow
fluorescent peaks will indeed arise upon double resonance.
For simplicity, we  consider beams along the
$\pm$ z direction only. When the atoms enter the
molasses region,  one half of the atoms of
any velocity $v$ are in the $ F = 2$ level and the other half
in the $F = 3$ level.
Within the molasses,
there is a redistribution of atoms between the two
ground states due to the laser fields. The rate at
which the cooling beam transfers atoms from  $F=3$ to
$F=2$ through the upper level $F'=3$ is given by\\
\begin{equation}
\eta_{32} = \Gamma_{3'3} \frac{\frac{I_c}{I_{s33'}}}{\sqrt{1 + 2(I_c/I_{s33'}) + 4(\Delta_{3'3}/\Gamma_{3'3})^2}} \frac{\Gamma_{3'2}}{(\Gamma_{3'3} + \Gamma_{3'2})}
\end{equation}
Here, $I_c$ = Intensity of the cooling beam,
$\Gamma_{ij}$ = linewidth of the transitions between states i and j,
$I_{sij}$= saturation intensity of transition from i to j,
$\Delta_{3'3}$ = $\Delta_c$ + $2\pi 121 MHz$ ,
$\Delta_c$ = $\delta_c$ -$\vec{k}.\vec{v}$.
Similarly, the rate at which the repumper laser transfers atoms from F= 2 to
F= 3 through F'=3 is
\begin{equation}
\eta_{23} = \Gamma_{3'2} \frac{\frac{I_r}{I_{s23'}}}{\sqrt{1 + 2(I_r/I_{s23'}) + 4(\Delta_{r}/\Gamma_{3'2})^2}}\frac{\Gamma_{3'3}}{(\Gamma_{3'3} + \Gamma_{3'2})}
\end{equation}
The symbols have meanings as given earlier, $\Delta_r = \delta_r - \vec{k}.\vec{v}$.\\
\indent Let us assume that there are N atoms with
velocity $v$. Of these, at a time t, $N_3$ are taken to be in level F = 3 and
$N - N_3$  in level F=2, neglecting any small number that may be  in the upper levels
F'=3' and 4'. $N_3$ should satisfy the  rate equation -
\begin{eqnarray}
\frac{dN_3}{dt} = 
&                 -(\eta_{32} + \eta_{23})N_3 + \eta_{23}N
\end{eqnarray}
The solution to this  is
\begin{equation}
N_3(t) = [ N_{30} - \frac{\eta_{23}}{\eta_{23} + \eta_{32}}N]e^{-(\eta_{23}+\eta_{32})t} + \frac{\eta_{23}}{\eta_{23} + \eta_{32}}N
\end{equation}
We see that the first term  decays at a rate
$\tau = \frac{1}{\eta_{32} + \eta_{23}}$ and the steady state value of the
population in level F=3 is given by the second term.
At double resonance \footnote{If the cooling laser is seen at resonance
with the cooling transition, it is then 121MHz away from the $3-> 3'$ transition},$\eta_{23} >> \eta_{32}$,
$N_3$ rises rapidly from $N_{30}$ to nearly N with a rise-time of a few tens of
microseconds, and the fluorescence is maximised.
Away from double resonance, the fluorescence is much smaller,
 as has been estimated using equation (4) and as is shown in Fig. 4b.
The fluorescence is maximum
when $|\delta_r| = \delta_c$ with a halfwidth of about 25 MHz
that  is governed by the velocity range of the atoms which
contribute to the fluorescence effectively.
This simple rate analysis gives a plausible explanation
of a narrow fluorescent peaks despite the presence of a thermal
velocity distribution. A rigourous density matrix calculation is given in a companion paper. 
\section{Conclusions}
Fluorescence from an optical molasses in $^{85}Rb$ in the presence of a 
pump (cooling) and probe (repumper) laser shows splitting of peaks, 
with the separation increasing  linearly with the  detuning of 
the pump. A  simple model explains both the  peak positions 
and their narrow
linewidth as arising from velocity selection due to the requirement of simultaneous 
Doppler resonance. It also provides a means of selecting out atoms of a narrow velocity 
range from a hot gas. 
\section{Acknowledgements}

\indent It is a pleasure to thank M.S. Meena for building the electronics modules, the
workshop for fabricating numerous components and  Prof. N. Kumar for his constant
encouragement. We also thank Prof. A. Kastberg, Umea University, Sweden, for his unstinting help
during the construction of the experimental setup. We are grateful to 
Dr. C.S. Unnikrishnan , TIFR, India, for the loan of the femto-Watt detector. 

\end {document}